\documentclass[prb,aps,twocolumn,tightenlines,floatfix]{revtex4}
\usepackage{graphicx}
\begin{document}

\title{Ferroelectricity in ultra-thin perovskite films}
\author{Na Sai, Alexie M. Kolpak, and Andrew M. Rappe}
\affiliation{The Makineni Theoretical Laboratories, Department of
Chemistry, University of Pennsylvania, Philadelphia, PA 19104-6323, USA}

\date{\today}

\begin{abstract} 
We report studies of ferroelectricity in ultra-thin perovskite films
with realistic electrodes. The results reveal stable ferroelectric
states in thin films less than 10~\AA\ thick with polarization normal
to the surface.  Under short-circuit boundary conditions, the
screening effect of realistic electrodes and the influence of real
metal/oxide interfaces on thin film polarization are investigated. Our
studies indicate that metallic screening from the electrodes is
affected by the difference in work functions at oxide surfaces. We
demonstrate this effect in ferroelectric PbTiO$_3$ and
BaTiO$_3$ films. 
\end{abstract}

\pacs{77.80.-e,77.84.Dy,77.22.Ej}

\maketitle

The effect of size on thin-film ferroelectricity has been
known for a long time, but has not been completely understood.
Initial experiments and mean field calculations based on the Landau
theory suggested that below a critical correlation volume~\cite{Lines}
of electrical dipoles between 10-100~nm$^3$, ferroelectricity vanishes
due to intrinsic size effects.~\cite{Kretschmer,Zhong} For thin films
with the polar axis perpendicular to the surface, incomplete
compensation of surface charges creates a depolarizing field that has
been shown to further reduce the polarization
stability.~\cite{Batra,Mehta}

Recently, however, monodomain ferroelectric phases have been observed
in very thin films, below ten unit-cells
thick.~\cite{Tybell,Bune,Lichtensteiger} Furthermore, Fong {\em et
al.}\ showed that ferroelectric phases can be stable down to
$\sim$12~\AA\ (three unit cells) in PbTiO$_3$ films by forming
180$^\circ$ stripe domains, suggesting that no fundamental thickness
limit is imposed by the intrinsic size effect in thin
films.~\cite{Fong} This idea has been corroborated by {\em ab initio}
calculations carried out on perovskite films, which tell that no
critical thickness exists for polarization parallel to the
surface~\cite{Almahmoud} and that polarization perpendicular to the
surface can exist in films three unit-cells thick if the
depolarization field is artificially removed.~\cite{Ghosez,Meyer}

On the other hand, it has been found that the depolarization field
plays a dominant role in reducing polarization normal to the surface
and depressing ferroelectric transition temperatures in thin films. In
a continuum model by Batra {\em et al.}~\cite{Batra} which includes
the depolarization effect, a critical thickness of 100~\AA\ for
perovskite films was analytically derived, assuming a Thomas-Fermi
screening length of 1~\AA\ for the metal electrodes. Also, a recent
first-principles calculation revealed that BaTiO$_3$ films with
SrRuO$_3$ electrodes lose ferroelectricity below $\sim$24~\AA\ (6 unit
cells),\cite{Junquera} thus suggesting that a minimum thickness limit
exists for useful ferroelectric films. While indeed a minimum film
thickness must be influenced by the polarization of the ferroelectrics
and the screening length of the electrodes,~\cite{Dawber} it is not
yet clear whether the depolarizing field can ever be completely
removed by realistic electrodes on ultrathin films, nor how monodomain
thin film ferroelectricity is affected by the choice of electrodes and
by the interactions at the metal/oxide interface.

In this paper, we provide answers to these questions. In particular,
we investigate how the critical thickness varies in different systems
with realistic electrodes.  We find that even though ferroelectricity
is lost in BaTiO$_3$ thin films, polarization close to the bulk value
can be stabilized in PbTiO$_3$ thin films with thickness less than
10~\AA; the behavior in BaTiO$_3$ is therefore not universal to all
ferroelectric thin films. Furthermore, we show that inequivalent
ferroelectric/electrode interfaces can assist in stabilizing
ferroelectricity in thin films.

We apply density functional theory (DFT) calculations~\cite{DACAPO} to
ultrathin ferroelectric capacitors that are constructed of PbTiO$_3$
and BaTiO$_3$ films sandwiched between two conducting electrodes.  Two
electrodes commonly used in ferroelectric devices are studied:
platinum (Pt) and the metallic oxide SrRuO$_3$.  Short-circuit
boundary conditions are imposed by the periodic boundary conditions
and electrodes of sufficient thickness.

\begin{figure}[htbp]
\includegraphics[width=6cm]{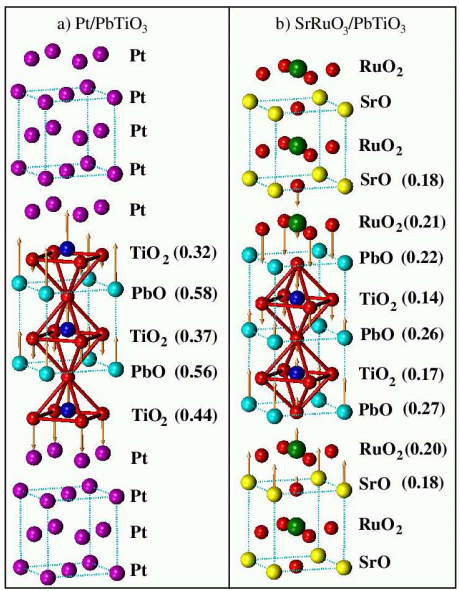}
\caption{(a) TiO$_2$-terminated Pt/PbTiO$_3$/Pt structure.  (b)
PbO-terminated SrRuO$_3$/PbTiO$_3$/SrRuO$_3$ structure. The
ferroelectric displacement patterns are shown by arrows.  The rumpling
of each layer (displacement of cations relative to anions) is marked,
in~\AA.}
\label{fig:structures}
\end{figure}

We consider AO (PbO or BaO) and TiO$_2$ ferroelectric terminations,
and we examine different ferroelectric-electrode interfaces, focusing
on the lowest energy one for each termination. Pt is most stable with
the first layer situated above the oxygen atoms on the TiO$_2$
terminated surface, and above A and O atoms on the AO-terminated
surface.~\cite{note-Pt} The periodically repeated supercells can thus
be described by the general formula~\cite{note-unitcell}
Pt$_4$/AO-(TiO$_2$-AO)$_m$/Pt$_5$ and
Pt$_4$/TiO$_2$-(AO-TiO$_2$)$_m$/Pt$_5$ with Pt electrodes and
(SrO-RuO$_2$)$_2$/AO-(TiO$_2$-AO)$_m$/(RuO$_2$-SrO)$_2$-RuO$_2$ and
(RuO$_2$-SrO)$_2$/TiO$_2$-(AO-TiO$_2$)$_m$/(SrO-RuO$_2$)$_2$-SrO with
SrRuO$_3$ electrodes (periodic boundary conditions means that the
ferroelectric slabs are separated by nine layers of electrode).
Figure~\ref{fig:structures} shows the structures for two
representative systems at $m=2$.  Although ferroelectric instabilities
can be present in the directions parallel and perpendicular to the
film, here we focus only on the latter situation, as our primary
interest is the depolarization effect.  Therefore, for all the
capacitors considered, the in-plane atomic positions are kept fixed at
the ideal perovskite positions and the in-plane lattice constants are
set equal to the experimental value for the corresponding bulk
ferroelectric perovskite, {\em i.e.}  $a$=3.935~\AA\ for PbTiO$_3$,
and $a$=3.991~\AA\ for BaTiO$_3$.~\cite{note-lattice}

We start by determining the fully relaxed structures of the
supercells.  The atomic positions and strains are fully relaxed along
the surface normal direction. Figure~\ref{fig:structures}(a) shows the
relaxed supercell structure of the TiO$_2$-terminated PbTiO$_3$ film
with Pt electrodes and $m$=2. The displacements from the
centrosymmetric structures are calculated. The relative displacement
between the cations and anions for each PbTiO$_3$ layer (see
Figure~\ref{fig:structures}(a)) clearly demonstrates that the
structure is in a ferroelectric state.  As Table~\ref{tab:pol} shows,
PbTiO$_3$ films with Pt electrodes are ferroelectric for all
thicknesses down to $m$=1.  For both terminations, the polarization
values in these systems are slightly larger than the corresponding
bulk value.~\cite{note-P}
\begin{table}
\caption{Polarization ($P$) and tetragonality ratio $c/a$ for
PbTiO$_3$ films with Pt and SrRuO$_3$ electrodes. The bulk values
are given in the last line. }
\begin{center}
\begin{tabular}{ccccc}
\hline\hline
termination &electrode &$m$  &$P$ (C/m$^2$) &$c/a$ \\\hline
PbO         &Pt        &1    &0.89          &1.109 \\
PbO         &Pt        &2    &1.00          &1.140 \\
PbO         &Pt        &4    &0.88          &1.058 \\
TiO$_2$     &Pt        &2    &0.86          &1.110 \\
TiO$_2$     &Pt        &4    &0.85          &1.055 \\
PbO         &SrRuO$_3$ &2    &0.36          &1.049 \\
TiO$_2$     &SrRuO$_3$ &2    &0.32          &1.040 \\\hline
            &          &bulk &0.75          &1.060 \\\hline\hline
\end{tabular}
\end{center}
\label{tab:pol}
\end{table}

These results suggest that bulk ferroelectric polarization can be
stabilized in thin films below 10~\AA\ with realistic electrodes. As
Figure~\ref{fig:structures} illustrates, an upward pointing
polarization leads to an enhancement of the displacements at the
bottom interface, and a reduction at the top interface, relative to
the interior layers.  This observation is in agreement with previous
DFT calculations with external fields.~\cite{Meyer} In going from
$m$=2 to $m$=4, both $P$ and $c/a$ decrease towards the bulk values as
the surface-to-volume ratio decreases and the surface effect is
averaged over more layers. To further check whether there exists a thickness
limit below which ferroelectricity disappears in this system, we
examine the structure at $m=1$.  We find that a single unit-cell has
a stable polarization of 0.89C/m$^2$, despite the stoichiometrically different
environment as compared to bulk PbTiO$_3$.\cite{Fong} We therefore find
strong evidence for an absence of a critical thickness for
ferroelectricity in PbTiO$_3$ films with Pt electrodes. In
contrast to the PbTiO$_3$ films, none of the BaTiO$_3$ structures were
found to be ferroelectric for $m$=2 or $m$=4, in agreement with the
results in Ref.~\onlinecite{Junquera}.

\begin{figure}[htbp]
\includegraphics[width=6cm]{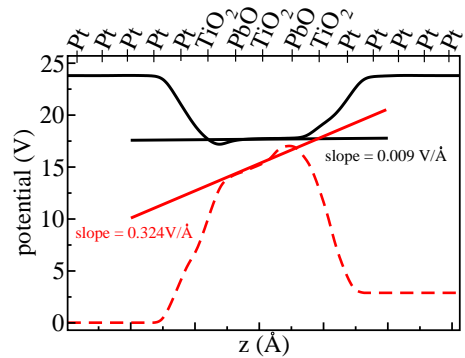}
\caption{Macroscopic-averaged  electrostatic potential along  the film
normal direction  of the TiO$_2$-terminated  Pt/PbTiO$_3$/Pt capacitor
(solid curve) and a free-standing  PbTiO$_3$ slab with a fixed bulk ferroelectric
displacement(dashed  curve). The  potential is constant inside the Pt
electrodes.}
\label{fig:pot}
\end{figure}

To make contact with earlier analytic theory,\cite{Batra,Dawber} we
examine the electric field in the Pt/PbTiO$_3$/Pt capacitor by
plotting the macroscopic-averaged~\cite{macro} electrostatic
potential, as shown in Figure~\ref{fig:pot}. Also shown in
Figure~\ref{fig:pot} is the potential in the free-standing PbTiO$_3$
film in which a bulk ferroelectric displacement perpendicular to the
surface is imposed.~\cite{note-freestanding} The slope of the
potential in the PbTiO$_3$ slab, which is the depolarizing field, is
significantly smaller in the capacitor (0.009~V/\AA), than in the
free-standing slab (0.324~V/\AA). Our calculations show that in the
absence of electrodes, the latter field brings the system back to the
paraelectric structure. The cancellation of a substantial fraction
(97\%) of the depolarizing field is due to metallic screening from the
grounded electrodes that compensates the polarization charge.

\begin{figure}[htbp]
\includegraphics[width=8cm]{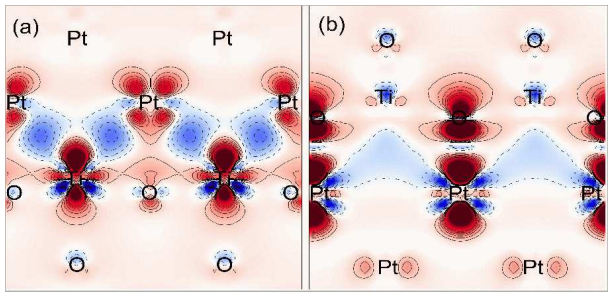}
\caption{Contour plot of the induced charge density for the
TiO$_2$-terminated Pt/PbTiO$_3$/Pt capacitor structure at the (a) top
interface (b) bottom interface.  (The Pt layers farther from the
interfaces are not shown.) Electron loss is given by solid lines and
electron gain by dotted lines. The induced charge density is found by
subtracting the charge densities of free-standing Pt and PbTiO$_3$
slabs from the total charge density in the capacitor.}
\label{fig:den}
\end{figure}

In addition, the screening is accompanied by the formation of unequal
local dipoles at the two interfaces due to different chemical bonding,
as shown in Figure~\ref{fig:den}.  On the top interface, the Pt and Ti
atoms lose charge, which is redistributed between the atoms, forming a
Pt-Ti alloy. On the bottom interface, where the Pt-O distance is the
shortest, the Pt and O atoms lose charge, the Pt from the $d_{z^2}$
orbitals, and the O from the $p_{z}$ orbitals,~\cite{Rao} while the Pt
$d_{xz}$ and $d_{yz}$ orbitals gain charge.  Similar behavior is
observed at the ferroelectric/electrode interfaces of the
AO-terminated capacitors. This inequivalent charge arrangement at the
two ferroelectric/metal interfaces is consistent with the different
interface polarizations we noted earlier.

With SrRuO$_3$ electrodes, the heterostructures also adopt a
ferroelectric state, as shown in
Figure~\ref{fig:structures}(b). However, unlike with Pt electrodes,
the polarization with SrRuO$_3$ electrodes is only half the bulk
value, indicating that the surface charges are only partially
compensated. The $c/a$ ratio for the SrRuO$_3$/PbTiO$_3$/SrRuO$_3$
capacitor falls to 1.045 as the film thickness decreases to two unit
cells.  The corresponding electrostatic potential shows a depolarizing
field of 0.07~V/\AA\ across the PbTiO$_3$ slab, which is much smaller
than in the free-standing PbTiO$_3$ film and explains why the
polarization is stabilized at a significant level.  Nevertheless, the
stronger field compared to that in the Pt/PbTiO$_3$/Pt structures also
confirms the weaker screening effect of SrRuO$_3$ compared to Pt and
other transition metals.~\cite{Dawber}

We note further that allowing the internal coordinates of the
SrRuO$_3$ electrodes, especially those in the boundary layers, to
relax together with the PbTiO$_3$ ions is crucial. When the SrRuO$_3$
ions are fixed in their ideal positions, we find only paraelectric
structures for either of the PbTiO$_3$ terminations. This result
indicates that the interaction between the PbTiO$_3$ and the SrRuO$_3$
electrodes plays a crucial role in stabilizing ferroelectricity: the
screening effect alone cannot account for the significant reduction of
the depolarizing field.

The results presented so far have suggested that ferroelectric
polarization normal to the surfaces can be stabilized in thin films
less than 10~\AA\ thick.  However they do not explain why the
ferroelectric instability can be retained in PbTiO$_3$ but not in
BaTiO$_3$ films when the same electrodes are applied. To address this
question, we turn now to look at the depolarizing field and screening
effects in these systems.  The depolarizing field was previously
addressed in phenomenological studies,~\cite{Batra2,Mehta,Dawber} and
is shown to scale with the spontaneous polarization of the
ferroelectrics, the screening length of the electrodes and the inverse
ferroelectric film thickness.  Hence the depolarizing field would be
expected to be negligible only in very thick films.

\begin{figure}[htbp]
\includegraphics[width=6cm]{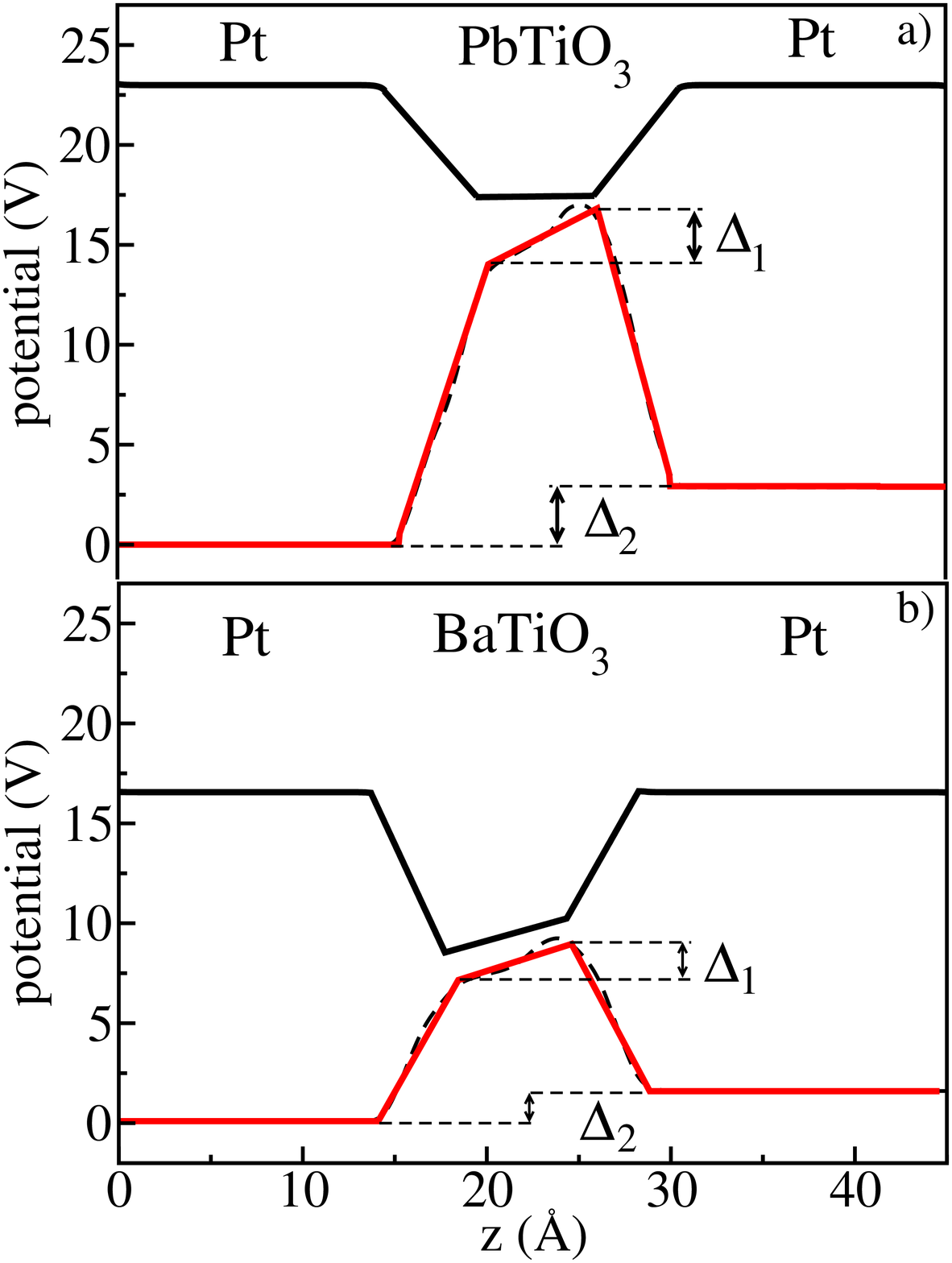}
\caption{ The electrostatic potential of a free-standing film with a
fixed bulk ferroelectric displacement (lower curve); the
self-consistent potential from solving the Poisson equation for
Thomas-Fermi screening charges (upper curve). Panel (a) is PbTiO$_3$
film and panel (b) is BaTiO$_3$ film.}
\label{fig:model}
\end{figure}

In our calculations, inspection of the electrostatic potential in the
free-standing film with a fixed bulk ferroelectric displacement shows
that the potential drop across the ferroelectric slab ($\Delta_1$) is
different from the potential difference between the two asymptotic
vacuum potentials ($\Delta_2$), as illustrated in
Figure~\ref{fig:model}.  This difference arises because the two
surfaces have different work functions, as a result of the
polarization orientation, parallel to the top surface normal and
anti-parallel to the bottom one.  Because Pauli repulsion keeps metal
electrons out of the ferroelectric, the potential drop that is
``seen'' and screened by the electrodes is $\Delta_2$, not $\Delta_1$.
An electrostatic analysis~\cite{Kolpak} of the potential and the
electric field in the slabs shows that this difference must be taken
into account when modeling the screening in all realistic systems.

We use a simple Thomas-Fermi model to highlight the effect of metallic
screening on a ferroelectric slab, without the complex structural and
electronic relaxations of DFT calculations.  We treat the DFT
macroaveraged electrostatic potential of the free-standing film as an
initial potential and find the metallic screening charge using
Thomas-Fermi theory.  We then find the total self-consistent potential
by solving the Poisson equation.~\cite{Kolpak} Figure~\ref{fig:model}
shows that in a PbTiO$_3$ slab of two unit cells, where
$\Delta_2/\Delta_1\sim1.1$, metallic screening from the electrodes
results in a potential that is close to constant in the PbTiO$_3$
slab. On the other hand, in a BaTiO$_3$ slab where
$\Delta_2/\Delta_1\sim0.9$, a similar calculation yields a screening
potential with a significantly larger depolarizing field.  These
Thomas-Fermi results can be explained as follows.  Suppose that the
metal screens a fraction $f$ of the potential drop $\Delta_2$.  For
$\Delta_2>\Delta_1$, this translates into screening a larger fraction
$f\Delta_2/\Delta_1$ of the ferroelectric potential drop $\Delta_1$.
However, if $\Delta_2<\Delta_1$, then the fraction of the
ferroelectric potential drop, $f\Delta_2/\Delta_1$, is less than $f$.
The difference in the polarization stability of ultrathin PbTiO$_3$
and BaTiO$_3$ films is therefore directly related to the relation of
$\Delta_2$ and $\Delta_1$ for the two materials.  We therefore
emphasize the importance of the inequivalence of the work functions in
determining the ferroelectric behavior of ultra-thin films at this
range of thickness.

In this paper, we have presented the first {\em ab initio}
demonstration of realistic electrodes stabilizing polarization normal
to the surface in ultra-thin films of $<$10~\AA\ thick. We have shown
that proper electrical and chemical boundary conditions are essential
in stabilizing ferroelectricity. Using various electrodes, we find
that monodomain ferroelectricity can persist down to one unit cell,
suggesting that thin film ferroelectrics are not specific to a single
system.  The ferroelectric polarization results in a difference in the
work functions at oxide surfaces that must be considered in modeling
the metallic screening.  We demonstrate this difference in PbTiO$_3$
and BaTiO$_3$ films in which we find significantly different screening
behaviors with the same electrodes.

The authors would like to thank I-Wei Chen and E. J. Mele for
insightful discussions.  This work was supported by the Office of
Naval Research, under Grants N00014-00-1-0372 and N00014-01-1-0365,
and by the NSF MRSEC Program, Grant DMR00-79909. Computational support
was provided by the HPCMO and by the DURIP program, as well as by the
NSF CRIF Program, Grant CHE-0121132. AMK is supported by an Arkema
Inc.\ fellowship.


\begin{thebibliography}{99}

\bibitem{Lines} M. E. Lines and A. M. Glass, {\em Principles and
Applications of Ferroelectrics and Related Materials} (Clarendon Press,
Oxford, 1977).

\bibitem{Kretschmer} R. Kretschmer and K. Binder, Phys. Rev. B. {\bf
20}, 1065 (1979).
 
\bibitem{Zhong} W. L. Zhong, Y.G. Wang, P.L. Zhang and B.D. Qu,
Phys. Rev. B.  {\bf 50}, 698 (1994).

\bibitem{Batra} I.P. Batra and B.D. Silverman, Solid State Comm. {\bf
11}, 291 (1972).
 
\bibitem{Mehta} R. R. Mehta, B. D. Silverman, and J. T. Jacobs,
J. Appl. Phys. {\bf 44}, 3379 (1973).

\bibitem{Bune} A. V. Bune. {\em et al.} Nature, {\bf 391}, 874 (1998).

\bibitem{Tybell} T. Tybell, C. H. Ahn, and J.-M. Triscone,
Appl. Phys. Lett. {\bf 75}, 856 (1999).

\bibitem{Lichtensteiger} C. Lichtensteiger, J.M. Triscone, J. Junquera
and P. Ghosez, Phys. Rev. Lett. {\bf 94}, 047603 (2005).

\bibitem{Fong} D. D. Fong {\em et al.}, Science {\bf 304}, 1650
  (2004).

\bibitem{Almahmoud} E. Almahmoud, Y. Navtsenya, I. Kornev, H.X. Fu and
L. Bellaiche, Phys. Rev. B. {\bf 70} 220102(R) (2004).

\bibitem{Ghosez} P. Ghosez and K.M. Rabe, Appl. Phys. Lett. {\bf 76},
2767 (2000).

\bibitem{Meyer}B. Meyer and D. Vanderbilt, Phys. Rev. B {\bf 63},
205426 (2001).

\bibitem{Junquera}J. Junquera and P. Ghosez, Nature {\bf 422}, 506
(2003).

\bibitem{Dawber} M. Dawber, P. Chandra, P.B. Littlewood and
J.F. Scott, J. Phys. Condens. Matter {\bf 15}, L393 (2003).

\bibitem{DACAPO} The calculations were performed within the
generalized gradient approximation to the DFT as implemented in the
DACAPO pacakge (http://www.fysik.dtu.dk/campos).  We have used the
ultrasoft pseudopotential (D. Vanderbilt, Phys. Rev. B {\bf 41}, 7892
(1990)).

\bibitem{note-Pt} The Pt electrodes can be oriented to form Pt~(100)
or Pt~(111) on the (100) surface of perovskites.  However, we choose
to focus on the Pt~(100) surface, as the orientation does not affect
the screening properties.

\bibitem{note-unitcell} The slabs all have access to a paraelectric
state with perfect inversion symmetry; this ensures that
ferroelectricity is due only to spontaneous symmetry breaking.

\bibitem{note-lattice} When slightly smaller in-plane lattice
constants of substrate materials such as SrTiO$_3$ are used instead,
all $c/a$ ratios are increased slightly compared to the values shown
in Table~\ref{tab:pol}.

\bibitem{note-P} Polarizations were computed by finding the ratio of
the thin film and bulk displacement amplitudes, then multiplying by
the bulk spontaneous polarization.

\bibitem{macro} The macroscopic-averaging (A. Baldereschi, S. Baroni
and R. Resta, Phys. Rev. Lett. {\bf 61}, 734 (1988)) is obtained by
convolving the planar-averaged potential (in $x$--$y$ plane) with a
window function in the $z$ direction using the PbTiO$_3$ and Pt
lattice constants for the window parameter. We select to focus on
$m$=2 in this analysis as it is the smallest structure that has a
center region one unit cell wide ($\approx$~4~\AA).

\bibitem{note-freestanding} This system can be realized by imposing
the atomic displacement of the bulk ferroelectric state in the center
of a slab, and relaxing the surface layers structure. The slope of the
potential is determined by the band gap and thickness of the
ferroelectric.

\bibitem{Rao} F. Y. Rao, M. Y. Kim, and A. J.  Freeman, Phy. Rev. B. {\bf
55}, 13953 (1997).

\bibitem{Batra2} I. P. Batra, P. Wurfel, and B. D. Silverman,
Phys. Rev. Lett. {\bf 30}, 384 (1973).

\bibitem{Kolpak} A. M. Kolpak, N. Sai and A. M. Rappe (unpublished).

\end{thebibliography}
\end{document}